\documentclass[aps,prd,twocolumn,nofootinbib,showpacs,superscriptaddress]{revtex4-1}

\usepackage{amssymb}
\usepackage{graphicx,graphics}
\usepackage{amsmath, amsthm}
\usepackage{epstopdf}
\usepackage{hyperref}

\newcommand{\be}{\begin{equation}}
\newcommand{\ee}{\end{equation}}

\begin{document}
\title{Understanding dynamical black hole apparent horizons}


\author{Valerio Faraoni}
\email{vfaraoni@ubishops.ca}
\affiliation{Physics Department and STAR Research Cluster, 
Bishop's University,\\
Sherbrooke, Qu\'ebec, Canada J1M~1Z7
}
\author{Angus Prain}
\email{A.Prain@hw.ac.uk}
\affiliation{School of Engineering and Physical Sciences,  Institute of Photonics and Quantum Sciences, Heriot-Watt University, Edinburgh EH14 4AS, UK}

\begin{abstract} 
\noindent
Dynamical, non-asymptotically flat black holes are best 
characterized by their apparent horizons. Cosmological black hole 
solutions of General Relativity exhibit two types of apparent 
horizon behaviours which, thus far, appeared to be completely 
disconnected. By taking the limit to General Relativity of a class 
of Brans-Dicke spacetimes, it is shown how one of these two 
behaviours is really a limit of the other.
\end{abstract}



\maketitle

\section{Introduction}
\label{sec:1}

The mechanics and thermodynamics of black holes developed in the 
1970s and culminating in the discovery of Hawking radiation rest 
on the concept of event horizon for stationary, 
asymptotically flat spacetimes. Correspondingly, thermodynamics 
was developed also for de Sitter space, which has a null horizon 
surface and is locally static below this horizon 
\cite{GibbonsHawking77}. However, realistic black holes are 
time-dependent and are not isolated: black holes are surrounded by 
astrophysical fluids, accretion disks, companions in binary 
systems, and are eventually embedded in the 
`background'\footnote{We use quotation marks because  
in general, due to the 
non-linearity of the field equations, any splitting of solutions 
into 
a background and deviations from it is not covariant.}  of the 
universe. As  a result, mass-energy is exchanged between a black 
hole and its surroundings and horizons are not stationary. Even 
Hawking radiation itself, corresponding to an energy outflow, causes the 
black hole mass to decrease and the horizon to shrink in a non-stationary manner suggesting that horizons are intrinsically non-stationary and proper event horizons are probably idealised abstractions of the real-world dynamical horizons. This 
backreaction effect is completely negligible for astrophysical 
black holes, but is theoretically important in the late stages of 
evaporation, in which the horizon changes dramatically.

The true black hole event horizon is defined as a connected component of 
the causal boundary of future null infinity (the cosmological event horizon is defined similarly) and, as such, it 
requires knowledge of the global spacetime structure to be located. 
In dynamical situations the event horizon so defined may not exist (this is 
the case of decelerated Friedmann-Lema\^itre-Robertson-Walker 
(FLRW) spaces). Nevertheless, such dynamical spacetimes still possess regions which locally behave like event horizons, trapping geodesics and cloaking singularities for example. In practice, it is these local versions of the horizon which are useful and meaningful in real astrophysical scenarios. 

One such astrophysical process is the production of gravitational waves from non-stationary gravitational sources.  During the last two decades 
the goal of theoretical research on gravitational waves has been 
the prediction of accurate waveforms for banks of templates used in 
the experiments designed to detect gravitational waves with giant 
laser interferometers such as {\em LIGO} and {\em VIRGO}. In these experiments, the 
signal-to-noise ratio is so low that signals can only be 
extracted by matching data with templates. The most promising 
processes 
for the direct detection of gravitational waves involve black holes 
and their horizons because gravitational collapse and merger 
processes produce clean signals. Numerical codes describing these 
processes cannot, by definition, know the entire future structure 
of spacetime;  event horizons, which are practically useless for 
numerical work, are {\em de facto} replaced by apparent and 
trapping horizons (see {\em e.g.}, \cite{numerical}). 

There is, 
therefore, a dichotomy in the 
theoretical community: while the more abstract mechanics and 
thermodynamics of black holes are mostly based on event horizons, 
which are null surfaces, the latter are replaced by apparent and 
trapping horizons in real world applications including gravitational wave physics. Apparent and 
trapping horizons are regarded with some suspicion by more 
mathematically oriented researchers because i) they depend on the 
foliation of spacetime (although they are coordinate-independent) 
\cite{WaldIyer91PRD}, and~ii) they are not null 
surfaces but 
can be spacelike, timelike, or change their causal character  
during the evolution of spacetime. Notwithstanding this {\em 
caveat}, there is little doubt that apparent horizons (AHs) are 
more useful than event horizons to locate `black hole boundaries' 
in dynamical and non-asymptotically flat spacetimes.  

Only a handful of exact solutions of General Relativity (GR) and of 
alternative theories of gravity are known which describe dynamical 
and/or non-asymptotically flat black holes, at least in some regions 
of the spacetime manifold. The known solutions are mostly spherically 
symmetric and describe inhomogeneities embedded in cosmological 
`backgrounds' (see Refs.~\cite{Galaxies, lastbook} for reviews).  
Contrary to the Kerr metric of GR, there are no uniqueness theorems 
that single out a particular solution of the Einstein (or 
alternative) field equations describing a central object surrounded 
by a cosmological or other non-trivial environment. Given this 
situation, to understand the dynamical behaviour of AHs  one must 
necessarily use particular exact  
solutions of the field equations. The zoo of known exact solutions 
with these features in GR and in alternative theories of gravity is 
quite small \cite{Galaxies, lastbook}. 

Apart from the well-known 
Schwarzschild-de Sitter-Kottler black hole, which is static in the 
region between the two (null, static) horizons, the first such 
solution of GR to be discovered was the McVittie metric 
\cite{McVittie}, which is spherically symmetric and describes a 
central inhomogeneity in a FLRW `background'
universe\footnote{See \cite{Krasinskibook} for a comprehensive 
review of inhomogeneous spacetimes.} and was introduced in order to 
study the effect of the cosmological expansion on local systems (a 
problem still debated today, see the recent review 
\cite{CarreraGiuliniRMD10}). The matter source for the McVittie 
metric 
is a fluid which has the character of a perfect fluid only far away 
from the central inhomogeneity. There is no flow of matter onto 
the central object. Studies of the causal structure and the 
AHs of the McVittie spacetime continue these days 
\cite{Nolan, AndresRoshina} and only 
recently has agreement been reached that the central object does 
indeed represent a black hole \cite{KaloperKlebanMartin10, 
LakeAbdelqader11, Roshina1, Roshina2, AndresRoshina, 
SilvaFontaniniGuariento12}.

A charged version of the McVittie metric has been proposed and 
studied \cite{ShahVaidya68, GaoZhang, FaraoniZambranoPrain}. More 
interesting are the so-called `generalized McVittie' spaces of GR 
sourced by an imperfect fluid with a radial spacelike heat flow 
onto the central object \cite{AudreyPRD}, the AHs of which were 
studied in \cite{GaoChenFaraoniShen08}. 
Recently, it 
was discovered that the McVittie geometry is also a  solution of 
cuscuton theory, a special representative of Ho\v{r}ava-Lifschitz 
gravity \cite{GuarientoAfshordi1}, while generalized McVittie 
metrics are 
also solutions of Horndeski theory \cite{GuarientoAfshordi2}. 

Another 
relevant solution of the Einstein equations is the 
Husain-Martinez-Nu\~nez one \cite{HusainMartinezNunez} which is 
sourced by a 
massless free scalar field $\phi$ and represents a black hole in 
a FLRW universe for part of the cosmic time. This solution was 
generalized by Fonarev \cite{Fonarev, HidekiFonarev} to the case 
of 
an exponential potential $V(\phi)=V_0 \exp(-\alpha \phi)$ for the 
scalar field. There are also Lema\^itre-Tolman-Bondi solutions of 
GR which represent dynamical black holes in 
cosmological `backgrounds': their AHs were studied 
in \cite{BoothBritsGonzalezVDB, GaoChenShenFaraoni11}. 

Still another 
solution of GR is the Sultana-Dyer black hole \cite{SultanaDyer04} 
sourced by a timelike and a null dust, which is conformal to the 
Schwarzschild metric and has been studied in order to shed light 
on the thermodynamics of AHs 
\cite{SaidaHaradaMaeda07, 
myHawkingT, Majhi}.  

Other spacetimes reported in the literature 
exhibit unphysical properties of their matter sources in at least 
some spacetime region \cite{McClureDyer}, or are not required to 
solve any field equations \cite{Lindesay}. Similar 
solutions of the field equations of alternative theories of gravity 
include the Brans-Dicke spacetimes found by Clifton, Mota, and 
Barrow \cite{CliftonMotaBarrow} and studied in \cite{climobarr}, the conformally transformed Husain-Martinez-Nu\~nez spacetime 
\cite{CliftonMotaBarrow} studied in \cite{FaraoniZambranocousin}, 
and the 
Clifton solution of $f(R)=R^{1+\delta}$ gravity 
\cite{CliftonCQG2006}, 
whose dynamical AHs were discussed in 
\cite{myClifton}.

\subsection{Fold-type or `C-curve' horizons}

\begin{figure}
\centering
\includegraphics[scale=0.45]{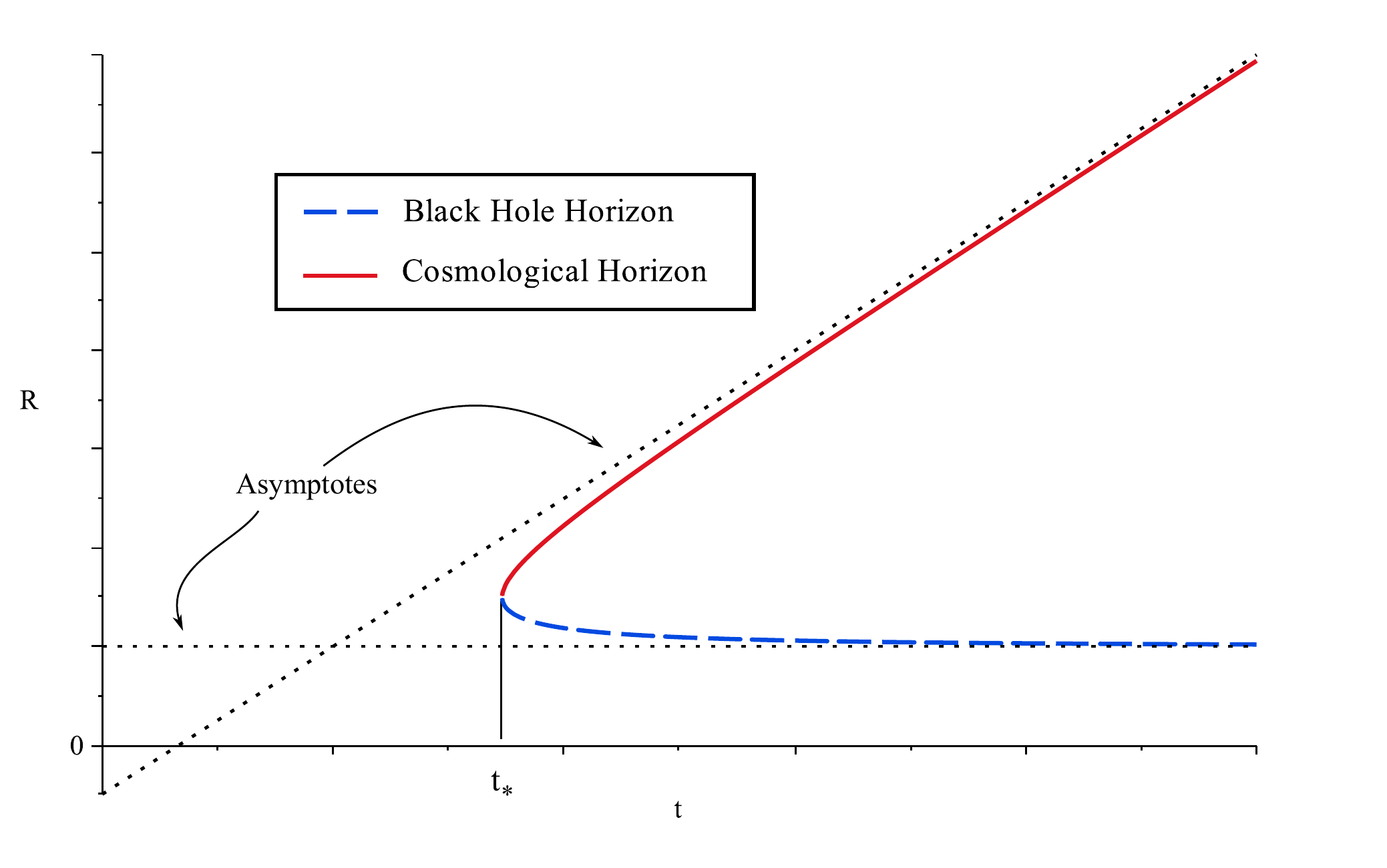}
\caption{Two apparent horizons appearing at time $t_*$, separating and asymptoting to what are intuitively conventional `cosmological' and `black hole' horizons.   \label{F:production} }
\end{figure}

\begin{figure} 
\centering
\includegraphics[scale=0.45]{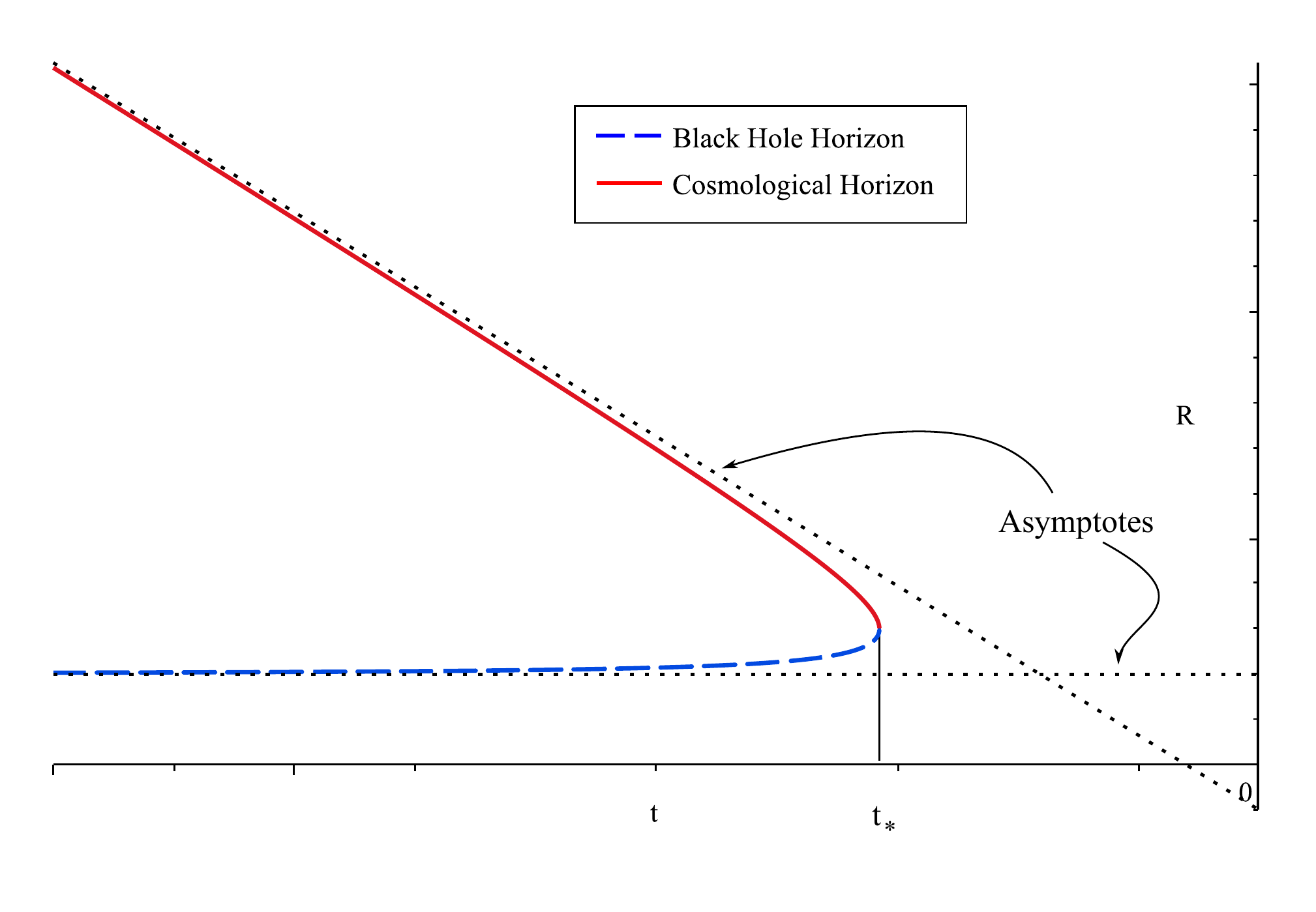} 
\caption{Two apparent horizons, one cosmological and one of black hole type, approaching a coincidence radius at time $t_*$, after which no apparent horizons are present. \label{fig:2} }
\end{figure}

The solutions mentioned above are all spherically symmetric, which 
allows for simplifications in the study of their structure and 
their AHs. In the presence of spherical symmetry, the 
AHs are located by the roots of the equation ({\em e.g.}, \cite{NielsenVisser})
\be
\nabla^cR\nabla_cR=0 \,,\label{E:condition}
\ee
where $R$ is the areal radius of spacetime. In general, this 
equation can only be solved numerically or, if its solutions can be 
expressed in analytical form, they are usually implicit. A rather 
common phenomenon encountered in the study of dynamical AHs 
\cite{lastbook} is the sudden appearance of a pair of 
AHs, or the merging and disappearance of pairs of AHs \cite{HusainMartinezNunez}. 
The two common phenomenologies are illustrated in Figs.~\ref{F:production} 
and \ref{fig:2}.

In Fig.~\ref{F:production}, a naked singularity present in an 
inhomogeneous universe gets covered by a black hole AH which 
appears together with a cosmological AH at a critical time $t_*$: the  
black hole AH starts shrinking while the cosmological AH expands. 
This type of phenomenology is encountered in the McVittie solution
\cite{Nolan, AndresRoshina, KaloperKlebanMartin10, 
LakeAbdelqader11, SilvaFontaniniGuariento12} 
and is interpreted in 
\cite{AndresRoshina}. 
Following the widely known example of the Schwarzschild-de 
Sitter-Kottler spacetime (which is a special case of McVittie for a static de Sitter `background'), for times $t$ less than the 
critical instant $t_*$ the black hole AH is larger than the cosmological AH and cannot be seen; at the critical time $t_*$ the two AHs coincide 
(and they are instantaneously null), or they are `created'; while 
for $t>t_*$ the black hole AH fits into the cosmological AH. The reversal of this, shown in Fig.~\ref{fig:2}, where the expansion is accelerating from an initially quasi-static state, the cosmological horizon shrinks and eventually engulfs the black hole horizon, leaving a naked singularity.  

\begin{figure} 
\centering
\includegraphics[scale=0.44]{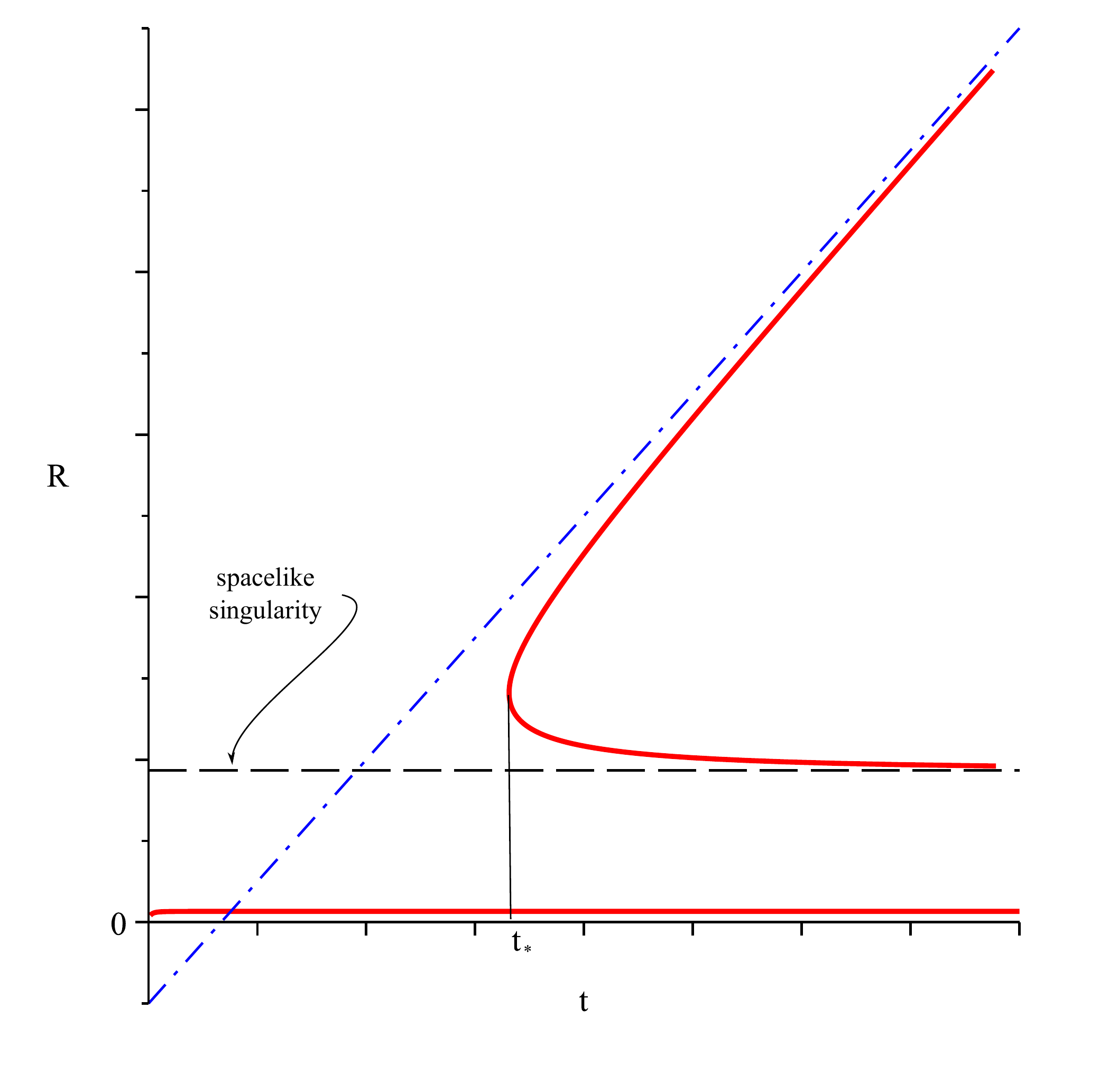}
\caption{The apparent horizon structure for the charged McVittie spacetime with $m=1$, $Q=1/2$, and a dust-dominated background $a(t)\propto t^{2/3}$.  \label{F:ccurve_normal} }
\end{figure}

\begin{figure} 
\centering
\includegraphics[scale=0.44]{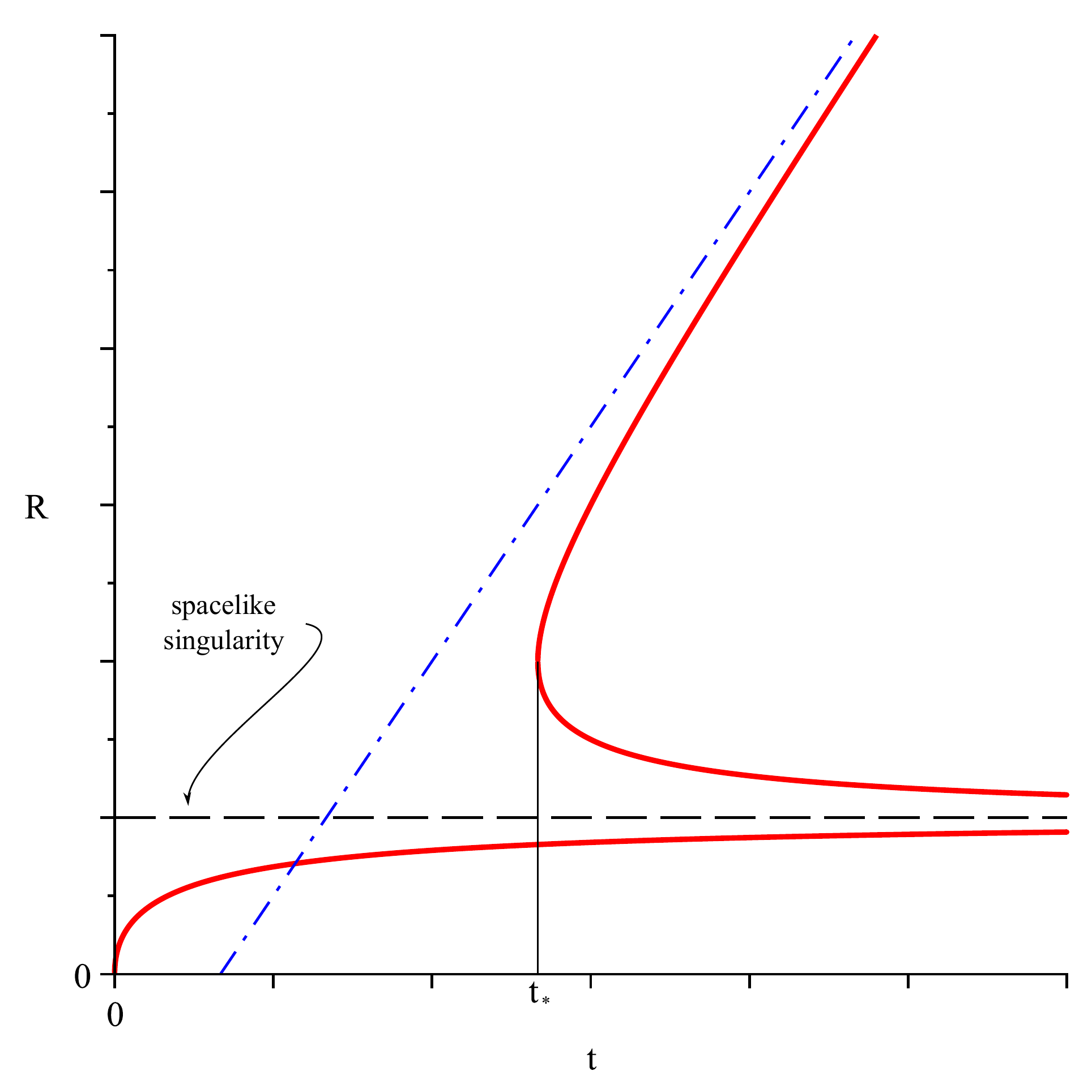}
\caption{The apparent horizon structure of the extremal McVittie spacetime $|Q|=m$, again in a dust-dominated background $a(t)\propto t^{2/3}$. Note that the inner and outer black hole horizons converge to the same radius at late times.   
\label{F:ccurve_from_below} }
\end{figure}

The 
picture corresponding to Fig.~\ref{F:production} for the actual exact McVittie solution at later times is richer than this simple scenario. In that case the 
black hole AH keeps shrinking and asymptotes to a spacelike spacetime 
singularity located at a finite areal radius. This singularity 
divides spacetime into two disconnected regions (but these two 
spacetime regions don't communicate---for all purposes they 
describe 
two different spacetimes).  In the charged version of McVittie spacetime, formally, a third
root exists in the spacetime region below the spacelike singularity.  Depending on the region of parameter space, this 
third root either asymptotes to a finite radius inside the spacelike singularity as shown in Fig.~\ref{F:ccurve_normal} 
or approaches this singularity from below, as shown in Fig.~\ref{F:ccurve_from_below}.
The asymptotic structure of the innermost pair of horizons in this charged case matches what one would expect from knowledge of the Reissner Nordstrom charged static black hole which possesses an inner and outer event horizon the locations of which coincide with the asymptotes of their counterparts in charged McVittie (including in the critical Reissner Nordstrom case).  After the Einstein-Straus Swiss-cheese 
model \cite{EinsteinStraus}, the McVittie solution of GR is probably the 
most famous spacetime describing a local object embedded in a 
universe and this is likely the first phenomenology of AHs that one 
is bound to encounter in a  survey of the literature \cite{Galaxies, 
lastbook}.

\begin{figure}
\centering
\includegraphics[scale=0.43]{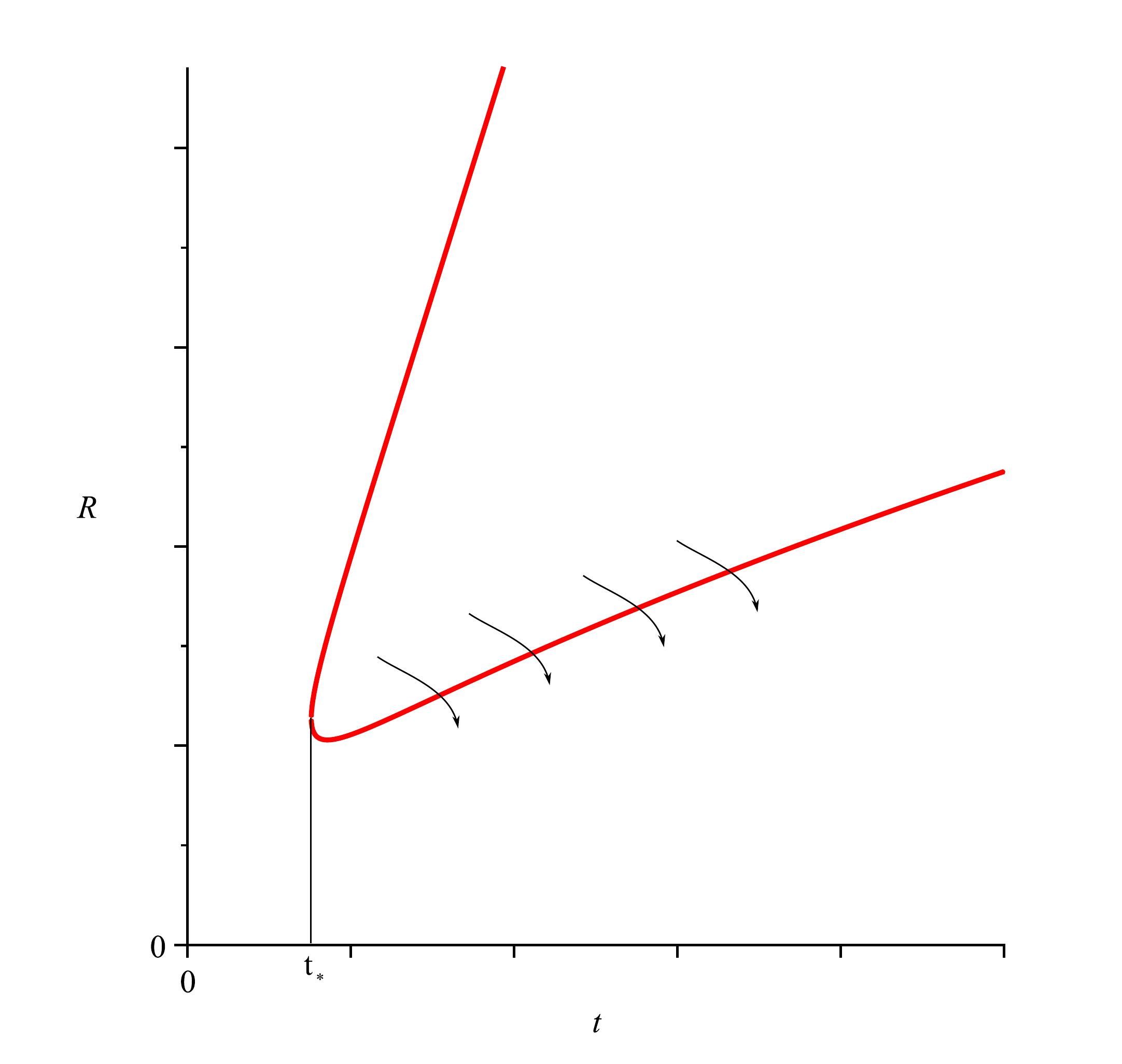}
\caption{The location of the apparent horizons for the comoving mass generalized 
McVittie solution with a dust-dominated scale factor $a(t)\propto t^{2/3}$. 
Here the inner (black hole) horizon grows as a positive flux of energy falls 
into the black hole and asymptotes to the spacetime singularity.   \label{F:generalized}}
\end{figure}

It should be pointed out that the McVittie and charged McVittie solutions are special in that they have zero radial flux of matter at the black hole horizon. Allowing for a radial energy flux into the central singularity, one is led to the so-called generalized McVittie solutions.
The generalized McVittie line element is given by \cite{AudreyPRD}
\begin{eqnarray}
ds^2 &=&  -\left( \frac{  1-\frac{M(t)}{2\bar{r}  a(t)} }{ 
1+\frac{M(t)}{2\bar{r}  a(t)} 
}\right)^2  dt^2  \nonumber\\
&&\nonumber\\
&\, & +  a^2(t) \left( 1+\frac{M(t)}{2 
\bar{r}  a(t)} 
\right)^4 \left( d\bar{r}^2+\bar{r}^2 d\Omega^2_{(2)} \right),
\label{generalizedMcVittie}
\end{eqnarray}
with $ M(t) \geq 0 $ an arbitrary  function of time 
and $  G^1_0 \neq 0$ (where $G_{ab}$ is the Einstein tensor), 
corresponding to a radial energy flow. The generalized McVittie spacetime has a singularity at $R=2M(t)$ (corresponding to $ \bar{r}=m/2$ in the notations of \cite{AudreyPRD}), 
where the energy density and the pressure diverge \cite{AudreyPRD}. 
The AH structure, as generically shown in Fig.~\ref{F:generalized}, is similar to the McVittie case but with an expanding black hole horizon as the black hole accretes energy and becomes more massive (see \cite{SilvaGuarientoMolina15}). However, when the surrounding matter is of phantom type, the black hole can lose mass and the black hole horizon can shrink, as discussed in  
\cite{GaoChenFaraoniShen08}. 

The phenomenology of AHs reported in Figs.~\ref{F:production}, \ref{F:ccurve_normal},  \ref{F:ccurve_from_below}, and \ref{F:generalized} will be referred to as fold-type or `C-curve' behaviour due to the shape of the curve. This phenomenology is reported also  for
Lema\^itre-Tolman-Bondi spacetimes \cite{BoothBritsGonzalezVDB, 
GaoChenShenFaraoni11}.

\subsection{Cusp-type or `S-curve' horizons}

\begin{figure} 
\centering
\includegraphics[scale=0.45]{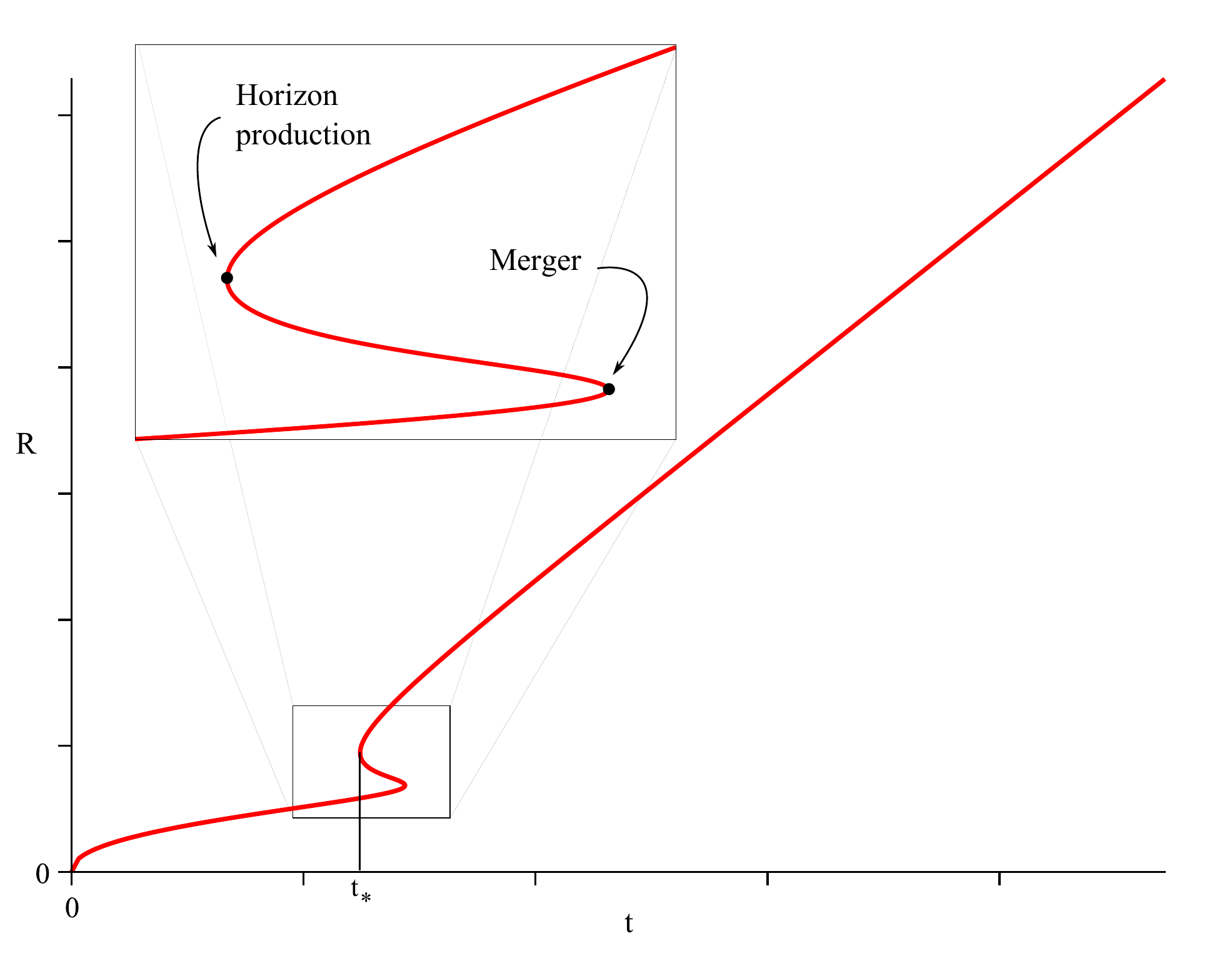}
\caption{The apparent horizon structure, which we call an `S-curve' or cusp-type, representing a pair of horizons appearing above an initially present horizon, one of which shrinks and merges with the first horizon, leaving behind a single cosmological horizon at late times. \label{F:scurve} }
\end{figure}

\begin{figure}
\centering
\includegraphics[scale=0.4]{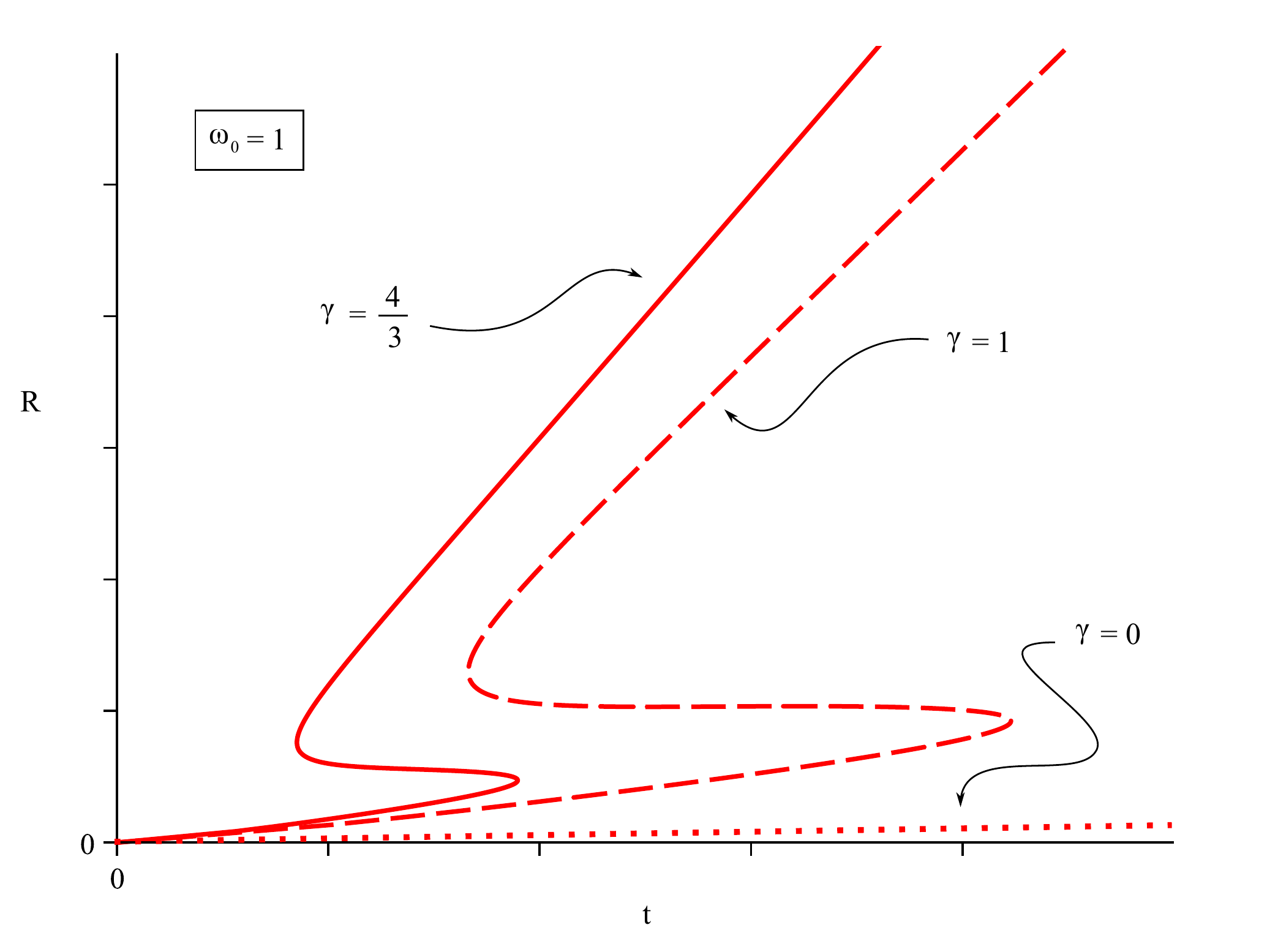}
\caption{The location of the AHs for a Clifton-Mota-Barrow exact solution to Brans-Dicke theory. Here $\omega_0$ is the Brans-Dicke parameter while $\gamma$ is an equation of state parameter with $\gamma=$ 0, 1, $4/3$ representing the solution sourced by a cosmological constant, dust and radiation respectively. \label{F:cusp}}
\end{figure}

A second type of phenomenological behaviour
of AHs of cosmological black holes is the one represented generally in 
Fig.~\ref{F:scurve}, which we will  refer to as cusp-type or `S-curve'. 
This terminology comes from catastrophe theory where curves of this shape describe what are called cusp-type catastophies. In this 
case, a single AH exists from the initial Big Bang, then two initially 
coincident AHs appear at time $t_*$. The larger 
(cosmological) AH expands forever, while the smaller (black hole) 
one shrinks, eventually meeting the smallest AH which in the 
meantime has been expanding. When these two black hole AHs 
(interpreted as inner and outer black hole AHs \cite{HusainMartinezNunez}) meet, they 
`annihilate' and disappear, leaving behind a naked central 
singularity in a  cosmological `background'. This S-curve 
behaviour was discovered by Husain, Martinez, and Nu\~nez in their
scalar field solution \cite{HusainMartinezNunez} and was 
subsequently reported 
\cite{myClifton} in the Clifton solution of 
metric $f(R)$ gravity \cite{CliftonCQG2006} and in the 
Clifton-Mota-Barrow 
family of solutions of Brans-Dicke theory \cite{CliftonMotaBarrow}
in a certain region of the parameter space \cite{climobarr}.  Indeed, Fig.~\ref{F:scurve} is precisely the AH structure for the Husain, Martinez, and Nu\~nez solution (with their parameter $\alpha=\sqrt{3}/2$ \cite{HusainMartinezNunez}). In Fig.~\ref{F:cusp} we also show the cusp-type behaviour of the AHs for an exact CMB solution which we will discuss more fully in the following section. 

Thus far, the s-curve and the c-curve phenomenologies of AHs seem 
completely disconnected and mutually exclusive---and what could the 
s-curve behaviour with three AHs have in common with spacetimes 
containing only two AHs and a singularity at a finite 
radius?   Here we will show that these two phenomenologies are not 
disconnected and that C-curve spacetimes correspond indeed to a 
limit of S-curve spacetimes when the lower 
bend in the S (where two AHs annihilate) is pushed to infinity. In
this limit a spacetime singularity appears at a finite radius, 
precisely at the location that would be occupied by the would-be 
critical AH resulting from the instantaneous merging of the inner 
and outer black hole AHs. This insight comes 
from the limit to GR of the Clifton-Mota-Barrow family of 
Brans-Dicke solutions.

\section{The Clifton-Mota-Barrow family of spacetimes}
\label{sec:2}
  
The Clifton-Mota-Barrow family (CMB) of solutions of Brans-Dicke theory 
is written as \cite{CliftonMotaBarrow}
\be\label{CMBmetric}
ds^2=-e^{\nu (\bar{r} )}dt^2+a^2(t) e^{\mu 
(\bar{r} )}\left(d\bar{r} ^2+\bar{r}^2d\Omega^2_{(2)}\right) 
\ee
in isotropic coordinates, where 
\begin{eqnarray} 
e^{\nu (\bar{r} )} & = &  
\left(\frac{1-\frac{m}{2\alpha \bar{r} }}{1+\frac{m}{2 \alpha \bar{r} 
}}
\right)^{2\alpha 
}\equiv A^{2\alpha} \,,\\
&&\nonumber\\
e^{\mu (\bar{r} )} & = & \left(1+\frac{m}{2\alpha \bar{r} }\right)^{4} 
A^{\frac{2}{\alpha}( \alpha-1)(\alpha +2)} \,, \\
&&\nonumber\\
\label{abeta}
a(t) & = & a_0\left(\frac{t}{t_0}\right)^{\frac{ 
2\omega_0(2-\gamma)+2}{3\omega_0\gamma(2-\gamma)+4}}\equiv 
a_{\ast}t^{\beta} \,,\\
&&\nonumber\\
\alpha & = & \sqrt{ \frac{ 2( \omega_0+2 )}{2\omega_0 +3} } 
\,,\label{7}\\
&&\nonumber\\
\rho^\text{(m)}(t, \bar{r} ) & = & \rho_0^\text{(m)} \left( \frac{ 
a_0}{a(t)} 
\right)^{3\gamma} A^{-2\alpha} \,, \label{density}
\end{eqnarray}
and the Brans-Dicke scalar field is given by \cite{CliftonMotaBarrow}
\be \label{scalart}
\phi(t, \bar{r} ) = \phi_0\left(\frac{t}{t_0} 
\right)^{\frac{2(4-3\gamma)}{  3\omega_0\gamma 
(2-\gamma)+4}}A^{-\frac{2}{\alpha }(\alpha^2-1)} \,.
\ee
Here $\omega_0$ is the Brans-Dicke 
parameter,  $m$ is a mass 
parameter, $\alpha, 
\phi_0, a_0$, $\rho^\text{(m)}_0$ and 
$t_0$ are positive constants ($\phi_0$, $\rho^\text{(m)}_0$, and 
$t_0$ are not independent). 
The matter source  is a perfect fluid with  energy density 
$\rho^\text{(m)}$, pressure  $P^\text{(m)}$, and equation of 
state $ P^\text{(m)}=\left( \gamma -1 \right) 
\rho^\text{(m)} \equiv w \rho^\text{(m)}$, where $\gamma$ is a 
constant 
\cite{CliftonMotaBarrow}. 

\begin{figure}
\centering
\includegraphics[scale=0.4]{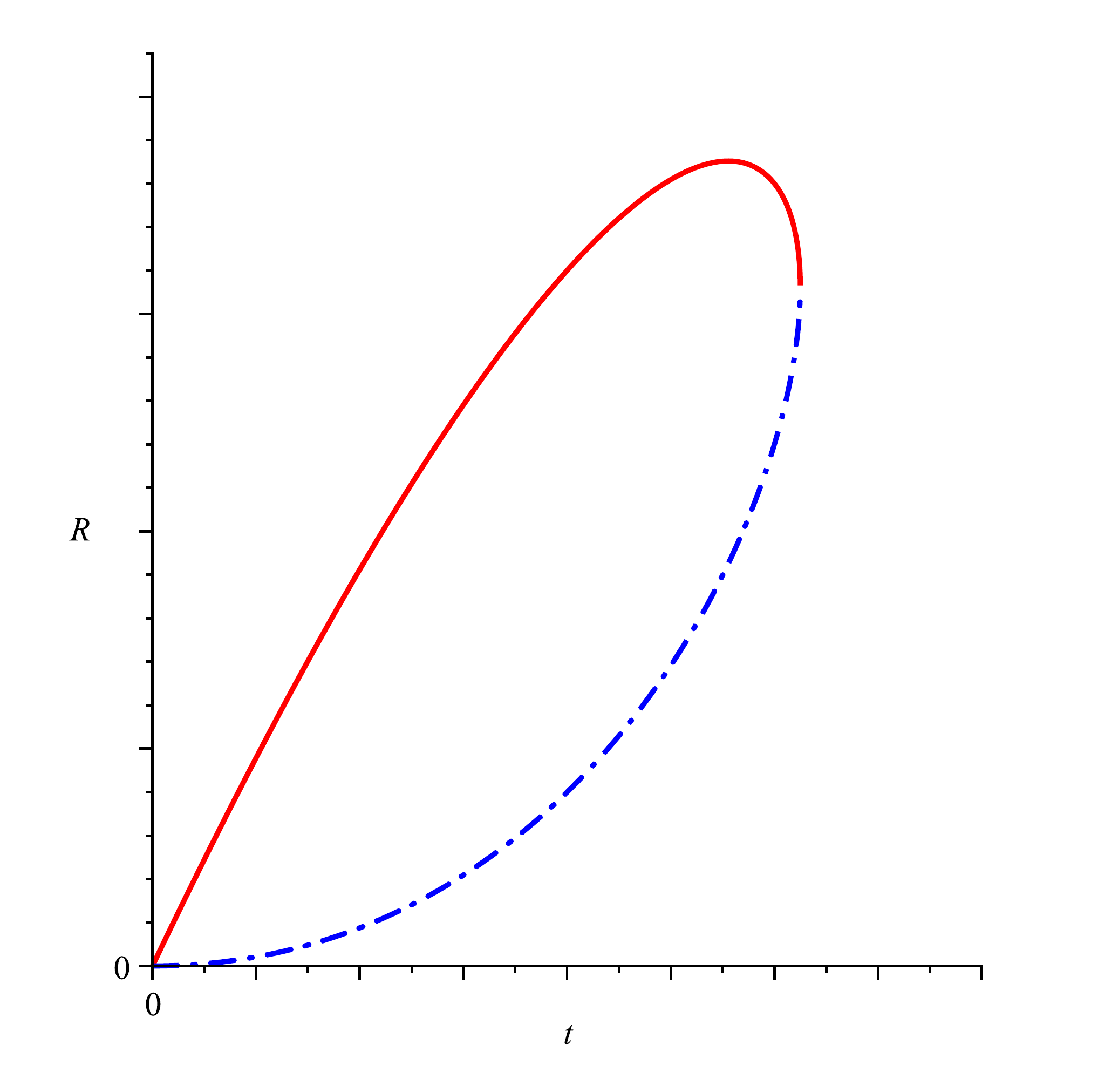}
\caption{A black hole and cosmological AH exist from the Big Bang, both expanding until the cosmological AH shrinks and eventually merges with the black hole AH after which there are no AHs and there exists a naked singularity at $R=0$.  \label{F:bubble}}
\end{figure}

The structure of the apparent horizons depends crucially on the equation of state parameter $\gamma$ and the Brans-Dicke parameter $\omega_0$. For completeness, when the expansion is positive, the condition \eqref{E:condition} reduces for this metric to 
\be
HR^2-\frac{(\alpha-1)(\alpha+2)}{\alpha^2}ma(t)A^{\frac{2(\alpha-1)(\alpha+1)}{\alpha}}-A^{\alpha+1}R=0
\ee
where $R$ is the areal radius. This condtion can have 0, 1, 2 or 3 roots for positive $t$ with the generic behaviour being that which is shown in Fig.~\ref{F:cusp} or alternatively a bubble like configuration as shown in Fig.~\ref{F:bubble}, which is a variant on the basic horizon merger structure which we described in Fig.~\ref{fig:2}. 

\subsection{The $\omega_0 \rightarrow \infty$ limit of the 
Clifton-Mota-Barrow family of spacetimes}

The Clifton-Mota-Barrow family of 
spacetimes was interpreted in 
Ref.~\cite{climobarr}, in which the limit to GR $\omega_0 
\rightarrow +\infty$ was briefly discussed, although the connection 
between S-curve and C-curve was missed there because of an incorrect 
figure (Fig.~4 of \cite{climobarr}).

The $\omega_0
\rightarrow +\infty $ limit of the line element (\ref{CMBmetric}) 
is quite 
interesting in itself: it is a late-time attractor of the 
generalized McVittie class of solutions (\ref{generalizedMcVittie}) 
of the Einstein equations. 
We have already remarked that there is no uniqueness theorem 
analogous to the Israel-Carter-Robinson theorems for 
non-isolated and/or dynamical black holes in GR. However, {\em within 
the restricted class of generalized McVittie solutions} \eqref{generalizedMcVittie}
\cite{AudreyPRD, GaoChenFaraoniShen08, GuarientoAfshordi1, 
GuarientoAfshordi2}, 
there is a unique late-time attractor \cite{FaraoniGaoChenShen09} 
and this is the only known such occurrence among classes of spacetimes  
describing cosmological black holes.

The $ \omega_0 \rightarrow +\infty$  limit of the line element is  
\begin{eqnarray} 
\label{grlm}
ds^2 & = & - \left( \frac{ 1-\frac{m}{2\bar{r} } }{ 
1+\frac{m}{2\bar{r} } 
}\right)^2 dt^2  \nonumber\\
&&\nonumber\\
&\, & +  
a^2(t) \left( 1+\frac{m}{2\bar{r} } \right)^4  \left( 
d\bar{r}^2+\bar{r}^2 d\Omega^2_{(2)} \right) \,,\\
&&\nonumber\\
a(t) & = & a_0 \left( \frac{t}{t_0} \right)^{ \frac{2}{3\gamma}} 
\,,\\
&& \nonumber\\
\rho^\text{(m)}(t) & = & \rho_0^\text{(m)} \left( \frac{t_0}{t} 
\right)^2 A^{-2}\,,
\end{eqnarray}
which is recognized as a special case of a generalized 
McVittie metric (\ref{generalizedMcVittie}).  The special case $ M(t)=M_0 
a(t) $, where $M_0$ is a constant, is the late-time  attractor 
of this class \cite{FaraoniGaoChenShen09} and is the 
$ \omega_0\rightarrow  \infty  $ limit of the 
Clifton-Mota-Barrow  spacetime~\eqref{CMBmetric}-\eqref{density}. The 
singularity of this attractor spacetime is located at the areal radius 
$R(t)=2M_0 a(t)$ and 
it expands comoving with the cosmic substratum.

The limit $\omega_0\rightarrow+\infty$ of the Clifton-Mota-Barrow 
spacetime has a peculiar feature: its two 
AHs are given by the simple expressions 
\cite{FaraoniGaoChenShen09}
\be\label{exactexplicitAHs}
R_\text{C, BH}=\frac{1}{2H} \left( 1\pm \sqrt{1-8m\dot{a}} \right)
\,,
\ee
where the subscripts C and BH stand for cosmological and black 
hole AHs, respectively, and an overdot denotes differentiation 
with respect to the comoving time $t$. It is very rare to be able 
to locate AHs 
analytically, and even rarer to do so explicitly \cite{lastbook}.  
Eq.~(\ref{exactexplicitAHs}) states that there are two and only 
two AHs provided that 
\be \label{condition}
1-8m\dot{a} >0 \,,
\ee
 where $a(t)=a_0 
t^\beta$, $\beta =\frac{2}{3\gamma} $ in the $\omega_0 \rightarrow 
+\infty$ limit, and 
$\dot{a}=\frac{a_0}{3\gamma} \, t^{\frac{2-3\gamma}{3\gamma}}$, otherwise there are none.

%
%
%

The case we will be interested in mainly is a dust-dominated universe which is described by the range $\gamma>2/3$.  In this case the exponent  $ \frac{2-3\gamma}{3\gamma} $ is negative which corresponds to the equation of state parameter $w>-1/3$ and hence
the universe is {\em decelerated} in this case. This will be the main case we are interested in because it is the case in which our equivalence between S- and C-curve horizons is most transparent. 

\begin{figure} 
\centering
\includegraphics[scale=0.4]{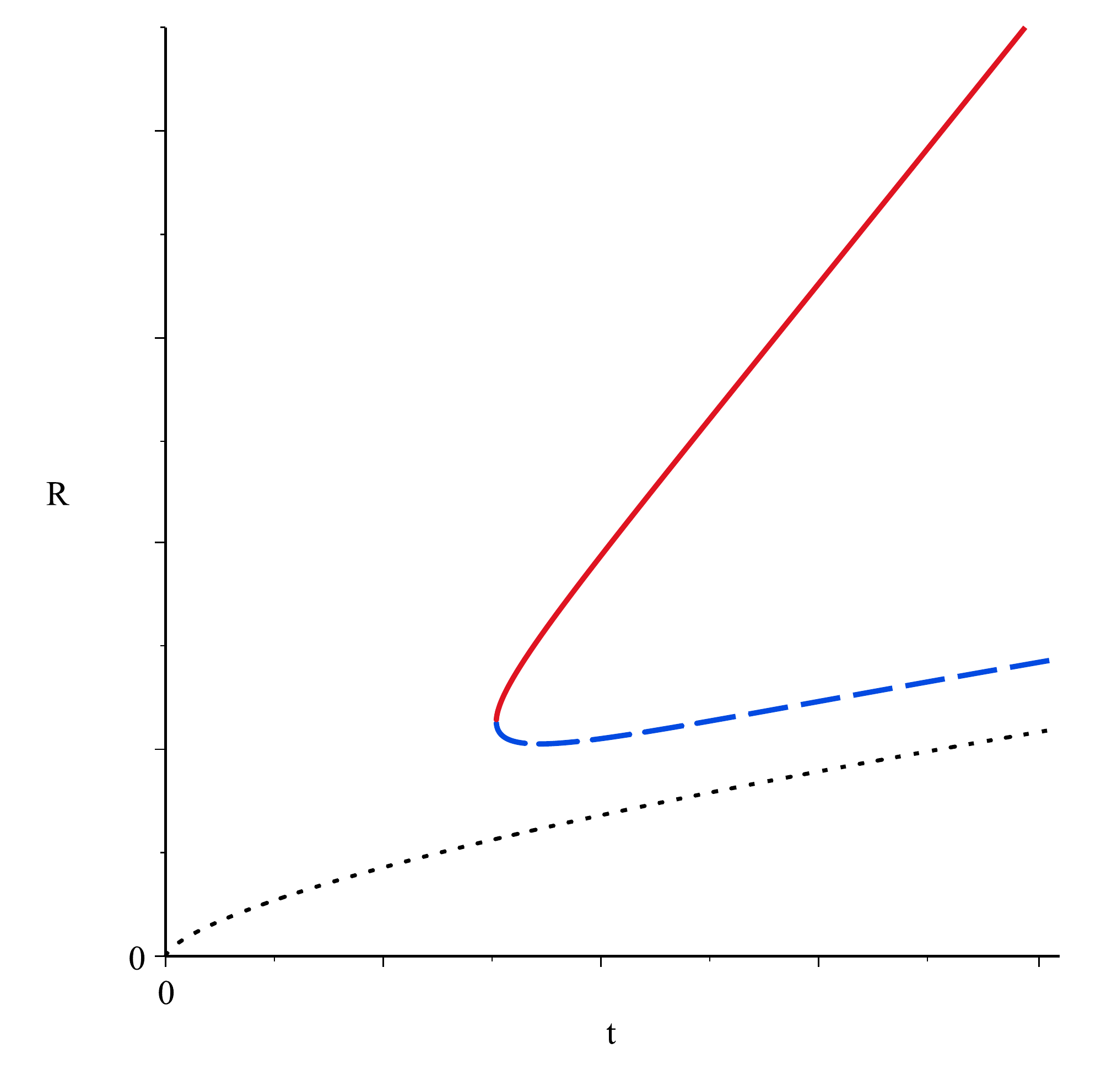}
\caption{Apparent horizons for the decelerated $\omega_0\rightarrow +\infty$ limit of the Clifton-Mota-Barrow solution 
for the fluid parameter value $\gamma=1$. \label{F:decelleration} }
\end{figure}

In this $\gamma>2/3$ case the two distinct AHs are present for
\be
t > t_{1*} = \left( \frac{16ma_0}{3\gamma} \right)^{\left| 
\frac{3\gamma}{2-3\gamma} \right|} \,.
\ee
These two AHs are created at $t_{1*}$ and subsequently have a C-curve 
behaviour as shown in Fig.~\ref{F:decelleration}. This is exactly as we showed schematically in Fig.~\ref{F:generalized} as an example of an accreting  cosmological black hole. In addition to the two apparent horizons there is a new finite-radius singularity indicated in Fig.~\ref{F:decelleration} by the black dotted line. 

Let us follow the evolution of the AHs of the $\gamma>2/3$ Clifton-Mota-Barrow 
spacetime as the Brans-Dicke parameter $ \omega_0$ becomes larger 
and larger.  This is limit is shown in the various S-curves\footnote{Fig.~4 of Ref.~\cite{climobarr} incorrectly 
depicts them as C-curves instead.} of 
Fig.~\ref{F:limiting} for increasing values of $\omega_0$: 
\begin{figure} 
\centering
\includegraphics[scale=0.47]{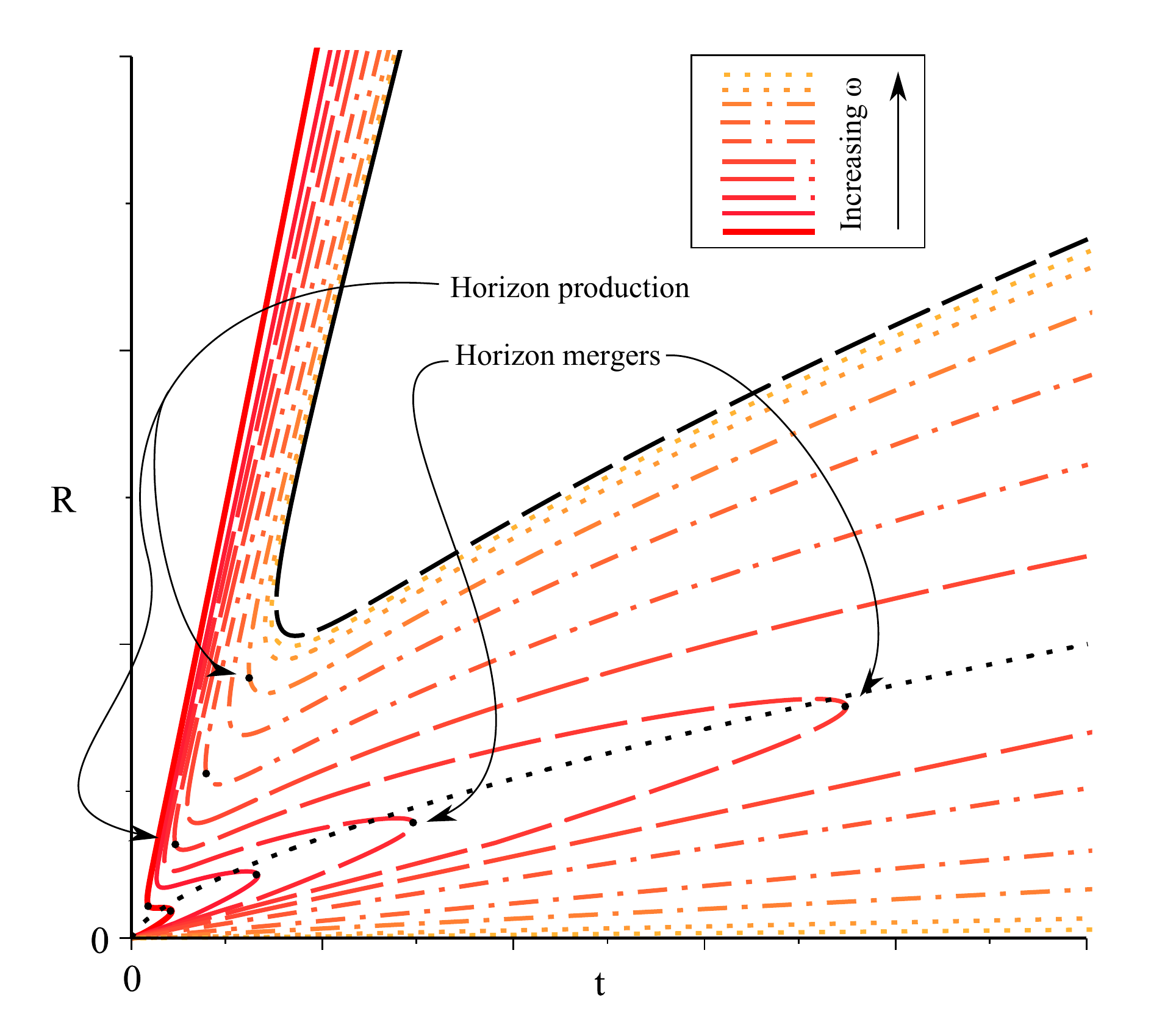}
\caption{The S-curves describing the apparent horizons of the Clifton-Mota-Barrow spacetime for $\gamma>2/3$ as   $\omega_0\rightarrow +\infty$. The lower bend of the S-curve is pushed to infinity as the parameter $\omega_0$ increases. We also plot the apparent horizons of the limit spacetime which in which the s-curve becomes a C-curve. The `new' singularity in the generalised McVitte limit spacetime is plotted in black dots. \label{F:limiting} }
\end{figure}
as $\omega_0$ grows, these 
curves become more and more 
stretched  and their lower bend (where the two innermost AHs merge and disappear) occurs at times $t_2$ which are 
larger and larger. As $\omega_0 \rightarrow +\infty$, also $t_2 
\rightarrow +\infty$ and the S-curve is broken, becoming a 
C-curve and the lowest branch of the S converges to 0 for all $t$. A spacetime singularity appears in this limit at a finite
areal radius, which is the location of the would-be `critical' 
horizon resulting from the merger of the  inner and outer black 
hole AHs occurring for finite values of $\omega_0$. 

The inner 
horizon (the 
$\omega_0 \rightarrow +\infty$ limit of the lower leg of the S-curve) is
usually not reported in the 
literature in plots describing the AHs structure of McVittie and 
generalized McVittie solutions because it belongs to a different 
spacetime disconnected from the previous one by the 
finite radius singularity. However, plotting the three AHs 
together clearly shows how the S-curve breaks into a C-curve in 
the limit. It looks as if AHs `want' to disappear in pairs and, 
by making only one of them survive in the $\omega_0\rightarrow 
+\infty$ limit, one somehow `splits' the spacetime instead. The 
mechanism 
and the deep reasons for this process are completely unclear, but 
one cannot help feeling that the structure of these AHs is telling  
us something about the spacetime itself.  

While the AHs momentarily coinciding at the lower bend 
of the S-curve are instantaneously null (with $dR_\text{AH}/dt 
\rightarrow \infty$ there), the black hole AH asymptoting to the 
finite-radius singularity in the generalized McVittie metric 
quickly becomes spacelike (with $dR_\text{AH} \rightarrow 0^{-}$). 
The 
AH in the disconnected spacetime ``below'' the finite 
radius singularity also becomes spacelike with $dR_\text{AH}/dt 
\rightarrow 0^{+}$.

\subsection{Other parameter ranges:  $0< \gamma < 2/3$ and $\gamma<0$}

When $0<\gamma<2/3$ (equivalent to $ -1<w<-1/3$ and corresponding 
to an {\em accelerated} universe) the exponent 
 $ \frac{2-3\gamma}{3\gamma} $ is positive, the 
inequality~(\ref{condition}) is satisfied, and there are two AHs for times less than a critical time $t_*$
\be
t\leq t_* \equiv \left( \frac{3\gamma}{16ma_0} 
\right)^{\frac{3\gamma}{2-3\gamma}} \,.
\ee
The two AHs exist during all times $0 < t \leq t_*$ approaching 
each other, then coincide and disappear at the time $t_*$, as we generically showed in Fig.~\ref{F:bubble}. 

One can interpret this behaviour as a black hole which is `swallowed' by the cosmological horizon as the latter shrinks due to the increasing acceleration of the expansion while for former grows as matter is accreted across it.\footnote{Due to a peculiarity of the power law, in fact accelerating spacetimes have a small $\dot{a}$ for early times and hence the cosmological horizon in that case always sits outside the black hole horizon, all the way back to $t=0$. On the other hand, decelerating spacetimes have a $\dot{a}$ which in fact diverges for small $t$, which is the reason why decelerated spacetimes begin their lives with a cosmological horizon which is much smaller than the black hole horizon. Effectively, decelerated spacetimes start off expanding very rapidly while accelerated spacetimes start off expanding mildly.} 

\begin{figure} 
\centering
\includegraphics[scale=0.4]{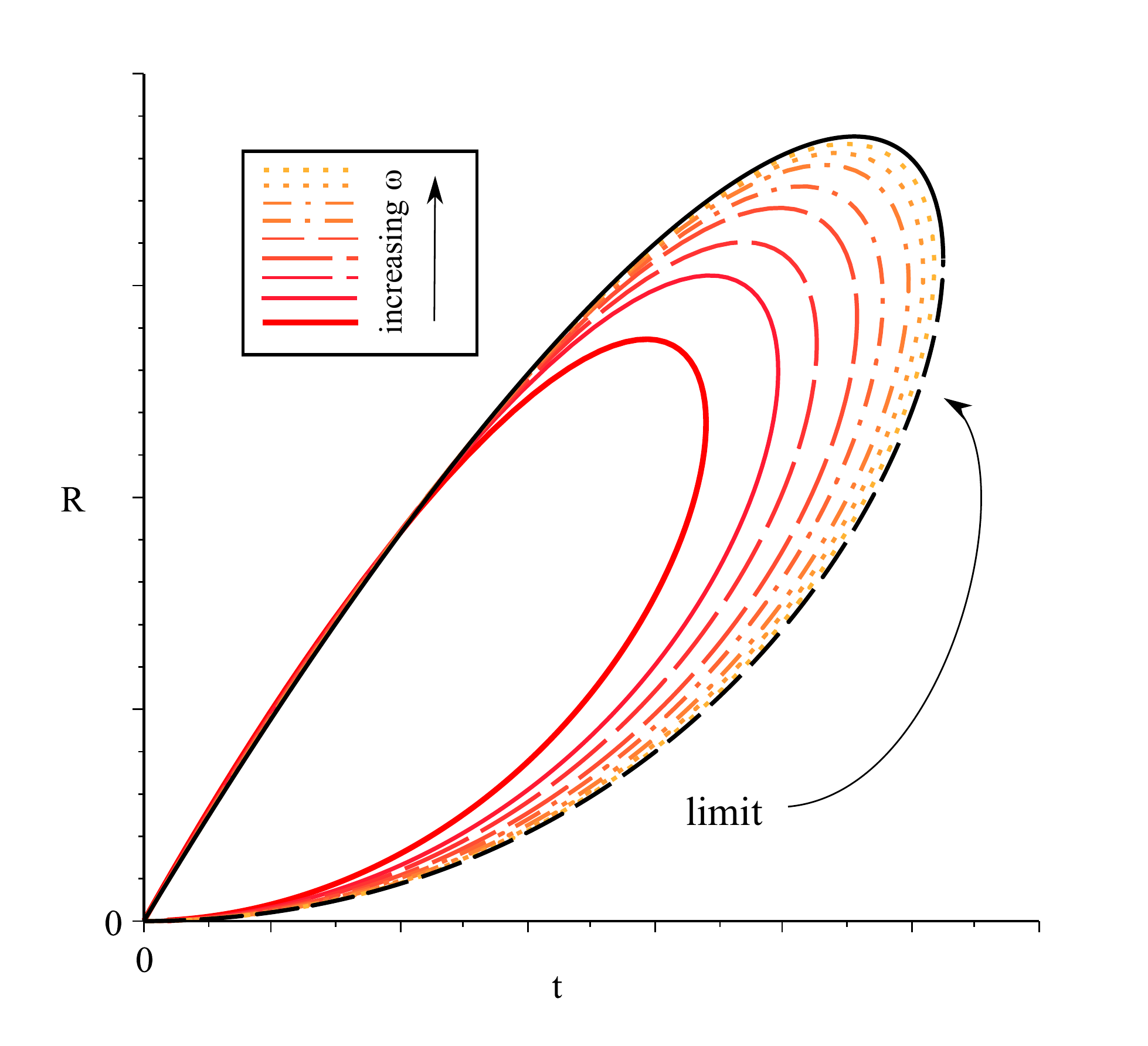} 
\caption{The apparent horizons in the $\omega_0\rightarrow +\infty$ limit of the Clifton-Mota-Barrow spacetime for $\gamma=1/3$. We also plot the appparent horizons of the limiting generalised McVitte spacetime\label{F:bubble_limiting}}
\end{figure}

Again we can follow the apparent horizon structure in the full CMB solution as the limit $\omega_0\rightarrow +\infty$ is carried out. In Fig.~\ref{F:bubble_limiting} we choose $\gamma=1/3$ and plot the horizons over time for increasing values of $\omega_0$.  The AHs form a closed curve 
touching the origin for each value of $\omega_0$ so the limit is less interesting and does not interpolate between cusp and fold structures.  

This accelerated expansion AHs are interpreted as a special case of the fold-type, with the fold pushed all the way back to the big bang singularity. 

\begin{figure}
\centering
\includegraphics[scale=0.4]{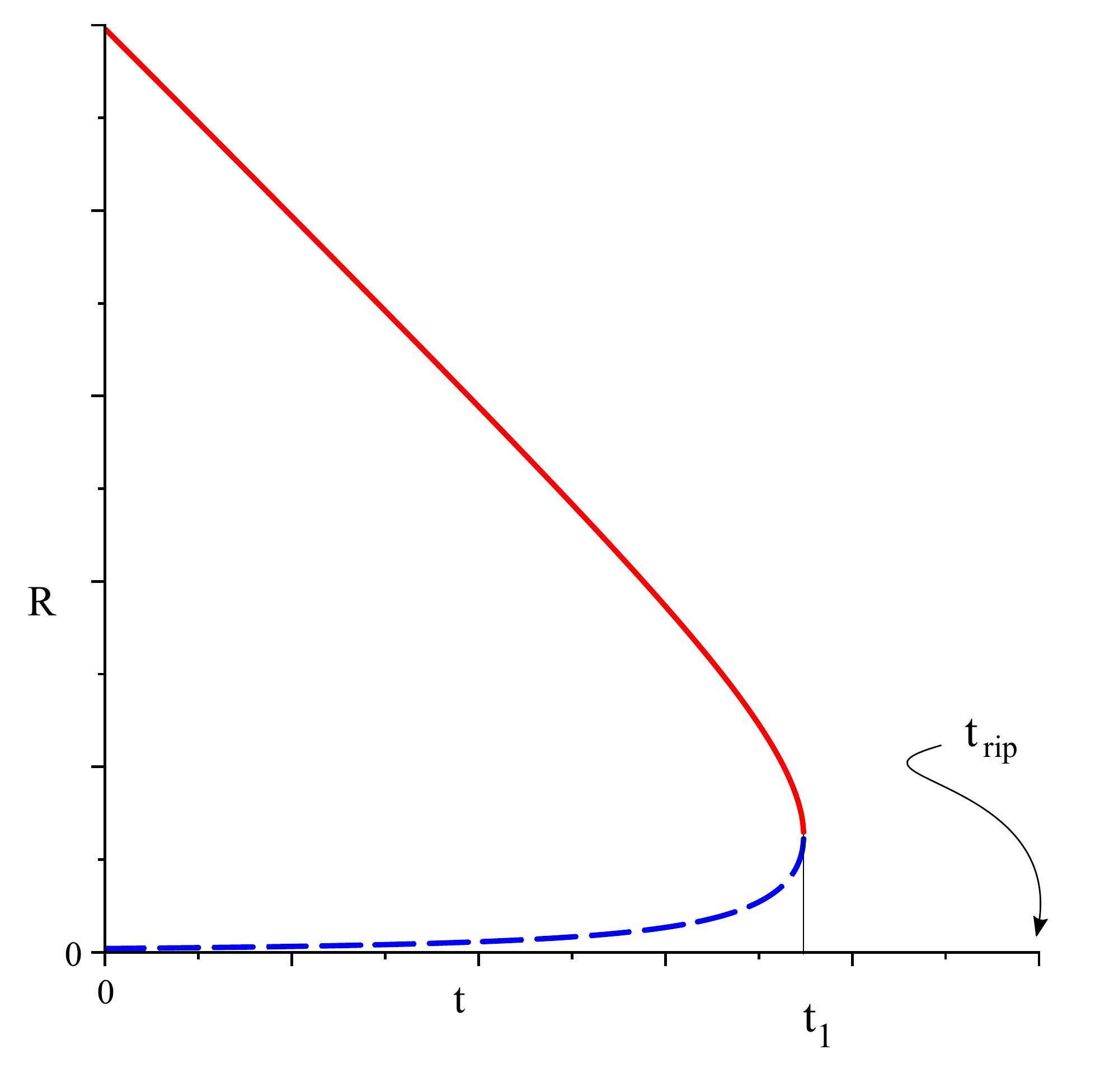}
\caption{The structure of the apparent horizons of the $\omega_0\rightarrow +\infty$ limit of the Clifton-Mota-Barrow solution for $\gamma=-1/3$ and $t_\text{rip}=10$. Note that the black hole apparent horizon asymptotes to $R=0$ in the past. \label{F:rip}}
\end{figure}

For completeness we mention the $\gamma<0$ spacetimes (corresponding to $w<-1$ and to a {\em phantom} 
universe with a big rip singularity at the time $t_\text{rip}$). The exponent $ \frac{2-3\gamma}{3\gamma} $ is again negative and the 
scale factor is
\be\label{formalscalefactor}
a(t)=\frac{a_0}{ \left( t_\text{rip}-t \right)^{\frac{2}{3|\gamma|} 
} } \,;
\ee
the inequality~(\ref{condition}) is satisfied, and there are 
two AHs, for times
\be
t < t_0 -\left( \frac{16ma_0}{3|\gamma|} \right)^{\frac{ 
3 |\gamma| }{2+3|\gamma |} } \equiv t_1 \,.
\ee
The two AHs coincide and disappear at $t_1$ as shown in Fig.~\ref{F:rip}. Note that, in our universe, phantom energy would not have dominated forever but would have started dominating only recently and the behaviour described
 by the formal solution~(\ref{formalscalefactor}) and shown for illustration 
 would have to be changed  accordingly. 


%
%
%
%
%
%

\section{Conclusions}
\label{sec:3}

Dynamical and non-asymptotically flat black holes are best 
characterized by their AHs. The known exact solutions of the 
Einstein equations which are spherically symmetric and represent 
dynamical black holes embedded in FLRW spaces exhibit two types of 
phenomenological behaviours, here dubbed ``S-curve'' and 
``C-curve'' types because of the shape of the AH 
radii versus comoving time. These phenomenologies show up also in 
(rare) cosmological black hole solutions of scalar-tensor and 
$f(R)$ gravity. Thus far, these two types of AHs 
behaviours appeared to be completely disconnected. However, by 
taking the limit to GR of the Clifton-Mota-Barrow spacetimes, one 
understands that the C-curve is just a limit of the S-curve as the 
lower bend of the S approaches infinity. Because no spacetime is 
known in which the AHs exhibit S-curve behaviour and can be  
located analytically and explicitly, one cannot control directly 
the location of this lower bend and adjust a parameter in order 
to push this bend to infinity. However, the limit $\omega_0 
\rightarrow +\infty$ of the Clifton-Mota-Barrow class of solutions 
of Brans-Dicke theory does exactly that for us: the C-curve is 
obtained essentially as the limit of S-curves as the lower bend of 
the S is pushed to infinity.  Therefore, this limit to GR connects
what thus far appeared to be two completely disconnected 
phenomenologies of AHs. What is more, AHs usually appear or 
disappear in pairs \cite{HusainMartinezNunez, 
Nolan, AndresRoshina} and 
inner/outer black hole AHs, in particular, seem to want to do 
the same: if only one horizon is selected through the limit to GR,
a spacetime singularity appears and the other AH is relegated to the 
associated (but disconnected) 
spacetime ``below the singularity''.  The reason why pairs of AHs 
can only be split by introducing a spacetime singularity between 
them remains a mystery.

\begin{acknowledgments} 

This research is supported by Bishop's University and by the 
Natural Sciences and Engineering Research Council of Canada ({\em 
NSERC}). 
\end{acknowledgments}



\end{document}